\documentclass[pdflatex,sn-nature]{sn-jnl}
 
\usepackage{graphicx}
\usepackage{multirow}
\usepackage{amsmath,amssymb,amsfonts}
\usepackage{xcolor}
\usepackage{booktabs}
\usepackage{hyperref}
\newcommand{\fesc}{f_{\rm esc}}
\newcommand{\fstar}{f_{\star,0}}
\newcommand{\xHI}{\bar{x}_{\rm HI}}
\newcommand{\ndot}{\dot{n}_{\rm ion}}
\newcommand{\tauobs}{\tau_e}
\newcommand{\xiion}{\xi_{\rm ion}}
\newcommand{\Msun}{M_\odot}
\newcommand{\Mpc}{{\rm Mpc}}
\renewcommand{\d}{\mathrm{d}}

\begin{document}

\title{The JWST early galaxy crisis resolved by a reionization degeneracy}

\author*[1]{\fnm{Zihan} \sur{Wang}}\email{zihan.wang@queens.ox.ac.uk}
\author*[2]{\fnm{Huanyuan} \sur{Shan}}\email{hyshan@shao.ac.cn}
 
\affil*[1]{\orgdiv{Department of Physics}, \orgname{University of Oxford}, \orgaddress{\city{Oxford}, \postcode{OX1 3PU}, \country{UK}}}
\affil*[2]{\orgdiv{Shanghai Astronomical Observatory}, \orgname{Chinese Academy of Sciences}, \orgaddress{\city{Shanghai}, \postcode{200030}, \country{China}}}
 
\date{\today}

\abstract{
JWST's discovery of unexpectedly bright $z>10$ galaxies has triggered claims that standard $\Lambda$CDM cannot reproduce their abundances, while estimates of the ionizing escape fraction $\fesc$ at $z>6$ have spanned a factor of four for over a decade. Here we show that both tensions arise from a structural degeneracy in reionization equations: global observables constrain only the product $\fesc\times \fstar$ (peak star formation efficiency), not individual parameters. We demonstrate that this degeneracy, previously considered a limitation, provides a precise diagnostic framework. By leveraging JWST UV luminosity function shapes to independently constrain $\fstar$, we derive robust bounds on $\fesc$. Joint profile-likelihood analysis across Gaussian, log-normal, and duty-cycle burst scatter models excludes the proposed crisis threshold ($\varepsilon > 3.5\%$) at $4.5\sigma$ confidence, with stochastic star formation histories strengthening rather than weakening the result. Combining these constraints with constant and evolving $\fstar$ measurements yields the first empirical reconstruction of $\fesc(z)$ across $z=7$–12. A constant-efficiency scenario ($\fesc \approx 10$–16\%) connects smoothly to low-redshift direct detections, whereas an evolving scenario ($f_{\rm esc} \approx 6\%$ at $z=12$) conflicts with low-metallicity ISM porosity expectations. JWST Cycle 3–4 will distinguish these pathways at $>2\sigma$, transforming a long-standing fundamental inference barrier into a powerful quantitative probe of early-universe physics.
}
\maketitle
\newpage

\section*{Introduction}
The discovery of unexpectedly UV-bright galaxies at $z > 10$ by JWST~\cite{Boylan-Kolchin:2022kae}
appeared to require baryon conversion efficiencies $\varepsilon \gtrsim 3.5\%$, far exceeding the
$\sim$1--2\% predicted by abundance-matching models and prompting claims of a crisis for
$\Lambda$CDM. Proposed solutions include enhanced star formation efficiency, feedback-free
starbursts, stochastic star formation histories~\cite{Mason:2022tiy}, and modified cosmologies,
yet no consensus has emerged. Independently, published estimates of the ionizing escape fraction
$\fesc$ at $z > 6$ span a factor of four, from $5\%$~\cite{Finkelstein:2019sbd} to
$21\%$~\cite{Naidu:2019gvi}, despite being derived from identical Planck and JWST
datasets~\cite{2015ApJ...802L..19R,Ma:2020vlo}. No systematic error has been identified, and
the disagreement has persisted for over a decade.

We show that these two apparently unrelated tensions share a single structural origin. The
reionization equations constrain only the product $\fesc \times \fstar$, where $\fstar$ is the
peak star formation efficiency, not the individual factors. Any combination preserving this product
fits all standard reionization observables identically. As shown in Fig.~\ref{fig:litridge},
published $\fesc$ estimates trace different positions along a one-dimensional degeneracy ridge in
the $(\fstar, \fesc)$ plane, and the JWST crisis corresponds to one extreme of this ridge where
$\varepsilon$ appears anomalously large. A companion analysis demonstrates that this degeneracy is
algebraically exact across all $\dot{n}_{\rm ion}$-dependent probes, confirmed with large suite of N-body
simulations~\cite{Wang:2026ljg}. Here we exploit the degeneracy as a measurement tool. JWST UV
luminosity function measurements constrain $\fstar(z)$ independently, breaking the ridge and
enabling the first empirical reconstruction of $\fesc(z)$ across $z = 7$--$12$.

We perform a joint profile-likelihood analysis over the $(\fstar, \fesc)$ parameter space, testing
three physically motivated scatter models: Gaussian, log-normal asymmetric, and a duty-cycle burst
model that captures the core physics of stochastic star formation histories. The crisis threshold
$\varepsilon > 3.5\%$ is excluded at $4.5\sigma$. Contrary to expectations, stochastic star
formation histories strengthen rather than weaken this exclusion, because the additional
model freedom benefits the baseline fit more than the crisis point. We combine the ridge constraint
with three independent JWST measurements of $\fstar(z)$ obtained from UV luminosity
functions~\cite{2024MNRAS.533.3222D}, the Uchuu-UniverseMachine
model~\cite{Prada2026}, and stellar mass functions~\cite{2024MNRAS.533.1808W}, and propagate all
uncertainties through a $5{,}000$-draw Monte Carlo.

The reconstruction yields two distinct physical pathways. A constant-efficiency scenario produces
$\fesc \approx 10$--$16\%$, connecting smoothly to direct Lyman-continuum detections at
$z < 4$~\cite{2018MNRAS.478.4851I, Steidel:2018wbo} and providing a parsimonious account of
escape fraction evolution across cosmic history. An evolving-efficiency scenario requires $\fesc$
to decline to $\sim 6\%$ by $z = 12$, in tension with the expected increase in interstellar medium
porosity at low metallicity~\cite{Ferrara:2022dqw}, and independently produces a Thomson optical
depth $\tau = 0.036$, rejected at $2.6\sigma$ by Planck. Three independent lines of evidence thus
disfavour the evolving pathway: the $\tau$ deficit, the contradiction with ISM porosity models, and
the convergence of two of three $\fstar(z)$ methods on the constant-efficiency reconstruction. For
over a decade, every attempt to measure how ionizing photons escaped the first galaxies produced a
different answer. Every answer was correct.

\section*{The degeneracy ridge}

The ionized fraction of the intergalactic medium evolves according to a balance between ionizing photon production and recombination (see Methods). The key structural property is that the reionization observables---Thomson optical depth $\tauobs$ and neutral fraction history $\xHI(z)$---depend on the galaxy population exclusively through the comoving ionizing emissivity $\ndot \propto \fesc \times \fstar \times \xiion$. Any combination of $\fesc$ and $\fstar$ that preserves this product yields identical reionization histories.

We calibrate the product using Planck 2018 cosmological parameters~\cite{Planck2018} and verify the degeneracy numerically at five parameter pairs along the ridge (Table~\ref{tab:ridge}). All five produce $\tauobs = 0.0551$, $\xHI(z=7) = 0.234$, and $\xHI(z=8) = 0.430$ to four significant figures, confirming that the degeneracy is exact. The calibrated product is $\fesc \times \fstar = 0.00247 \pm 0.00032$, with the $13\%$ error propagated from the Planck $\tauobs$ uncertainty. The baryon conversion efficiency ranges from $1.2\%$ to $4.9\%$ across the ridge, demonstrating that the crisis interpretation ($\varepsilon > 3.5\%$) depends entirely on which position is assumed but not on the data.

\begin{figure}
\centering
\includegraphics[width=\columnwidth]{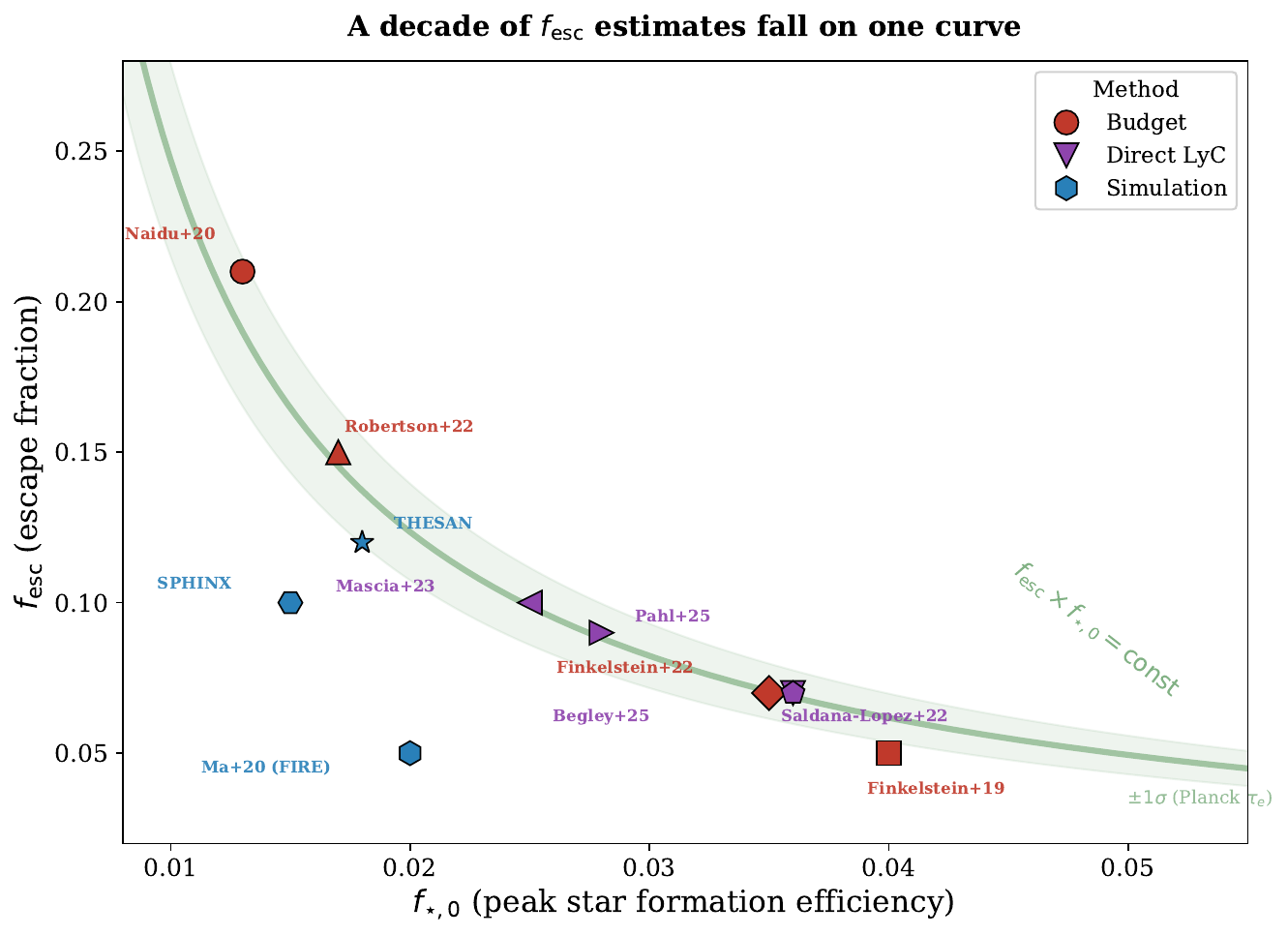}
\caption{A decade of published $\fesc$ estimates collapse onto the degeneracy ridge $\fesc \times \fstar = {\rm const}$ (green band: $\pm 1\sigma$ from Planck $\tauobs$). Red: photon budget models; purple: direct Lyman continuum; blue: RHD simulations. The factor-of-four scatter in $\fesc$ ($5$--$21\%$) traces different assumed $\fstar$ values, not genuinely discrepant physics.}
\label{fig:litridge}
\end{figure}

Figure~\ref{fig:litridge} shows that eleven published $\fesc$ estimates from the past decade---derived from photon budget models, direct Lyman continuum detections, and radiation-hydrodynamic simulations---all fall along or near the degeneracy ridge within their uncertainties. Values spanning $\fesc = 5\%$ to $21\%$ are not discrepant measurements; they are consistent determinations of the same product, differing only in the assumed $\fstar$.

\begin{table}[b]
\caption{Three solutions along the degeneracy ridge. All three produce identical $\tauobs$ and $\xHI(z)$ despite a factor-of-four range in $\varepsilon$.\label{tab:ridge}}
\begin{tabular}{lccc}
\toprule
Parameter & A (high $\fstar$) & B (baseline) & C (high $\fesc$) \\
\midrule
$\fstar$      & 0.049  & 0.019  & 0.012  \\
$\fesc$       & 0.05   & 0.13   & 0.20   \\
$\fesc \times \fstar$  & 0.00245 & 0.00247 & 0.00240 \\
$\varepsilon$ & 4.9\%  & 1.9\%  & 1.2\%  \\
$\tauobs$     & 0.0551 & 0.0551 & 0.0551 \\
$\xHI(z{=}7)$ & 0.234  & 0.234  & 0.234  \\
$\xHI(z{=}8)$ & 0.430  & 0.430  & 0.430  \\
\bottomrule
\end{tabular}
\end{table}

\section*{Breaking the degeneracy with JWST}

While reionization observables constrain only the product, the UV luminosity function constrains $\fstar$ independently through its shape. A higher $\fstar$ shifts the UV luminosity--halo mass relation, changing the predicted UVLF normalisation, faint-end slope, and bright-end cutoff. We quantify this with a joint profile likelihood over a $14 \times 16$ grid in $(\fstar, \fesc)$, profiling over the faint-end slope $\alpha_{\rm lo}$ and UV scatter $\sigma_{\rm UV}$ at each point (see Methods). The UVLF data comprise 28 points from Donnan et al.~\cite{2024MNRAS.533.3222D} at $z = 9$--$12.5$, supplemented by Harikane et al.~\cite{2025ApJ...980..138H}.

\begin{figure}
\centering
\includegraphics[width=\columnwidth]{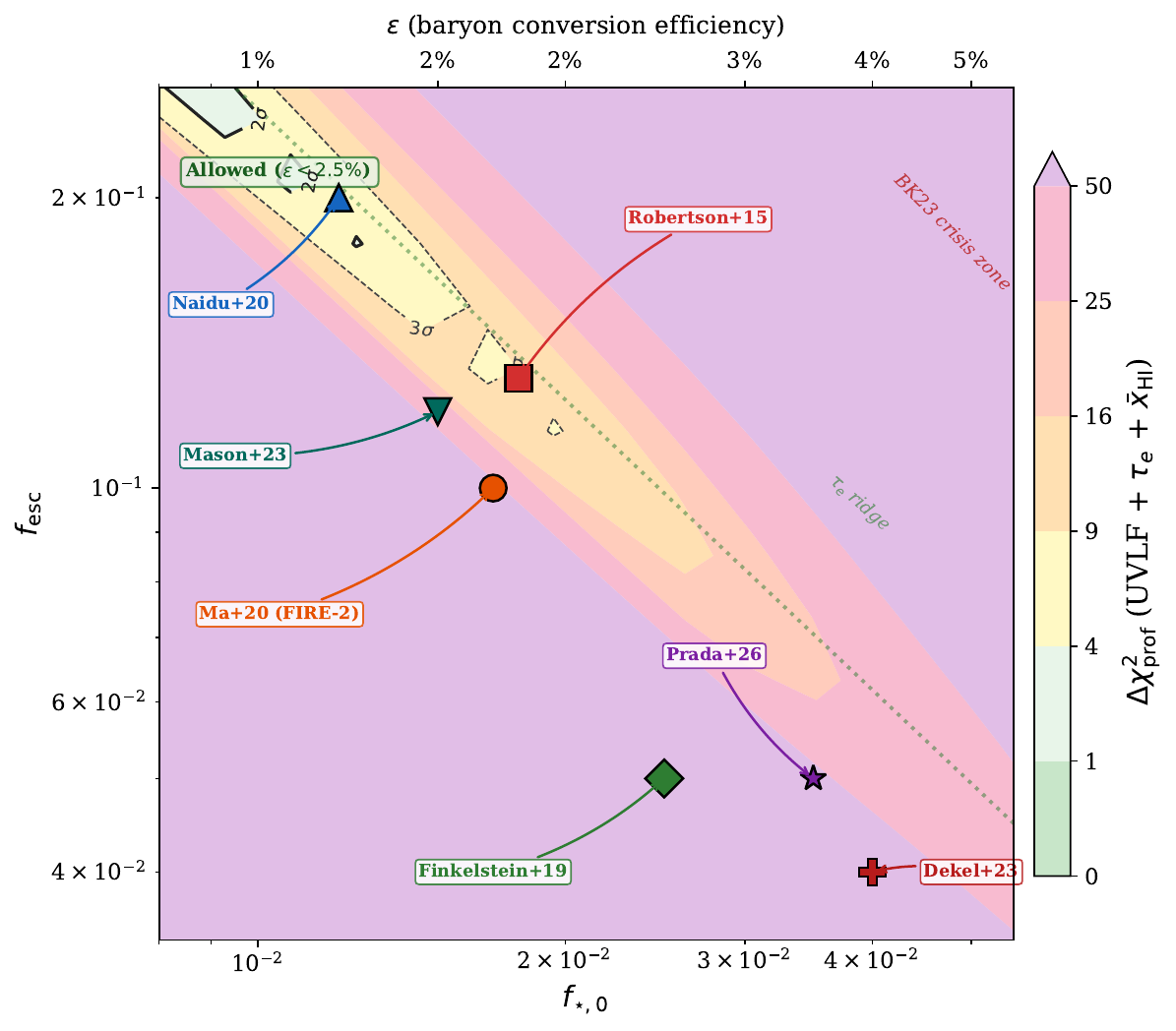}
\caption{Joint profile likelihood $\Delta\chi^2_{\rm prof}$ in the $(\fstar, \fesc)$ plane, with nuisance parameters profiled at each grid point. The green dotted line marks the degeneracy ridge. The UVLF shape breaks the degeneracy, truncating the allowed region at $\varepsilon \approx 2.5\%$ and excluding the BK23 crisis zone (red). Seven published solutions are overlaid; only those within the $2\sigma$ contour are consistent with the data.}
\label{fig:ridge}
\end{figure}

\begin{figure*}
\centering
\includegraphics[width=0.95\textwidth]{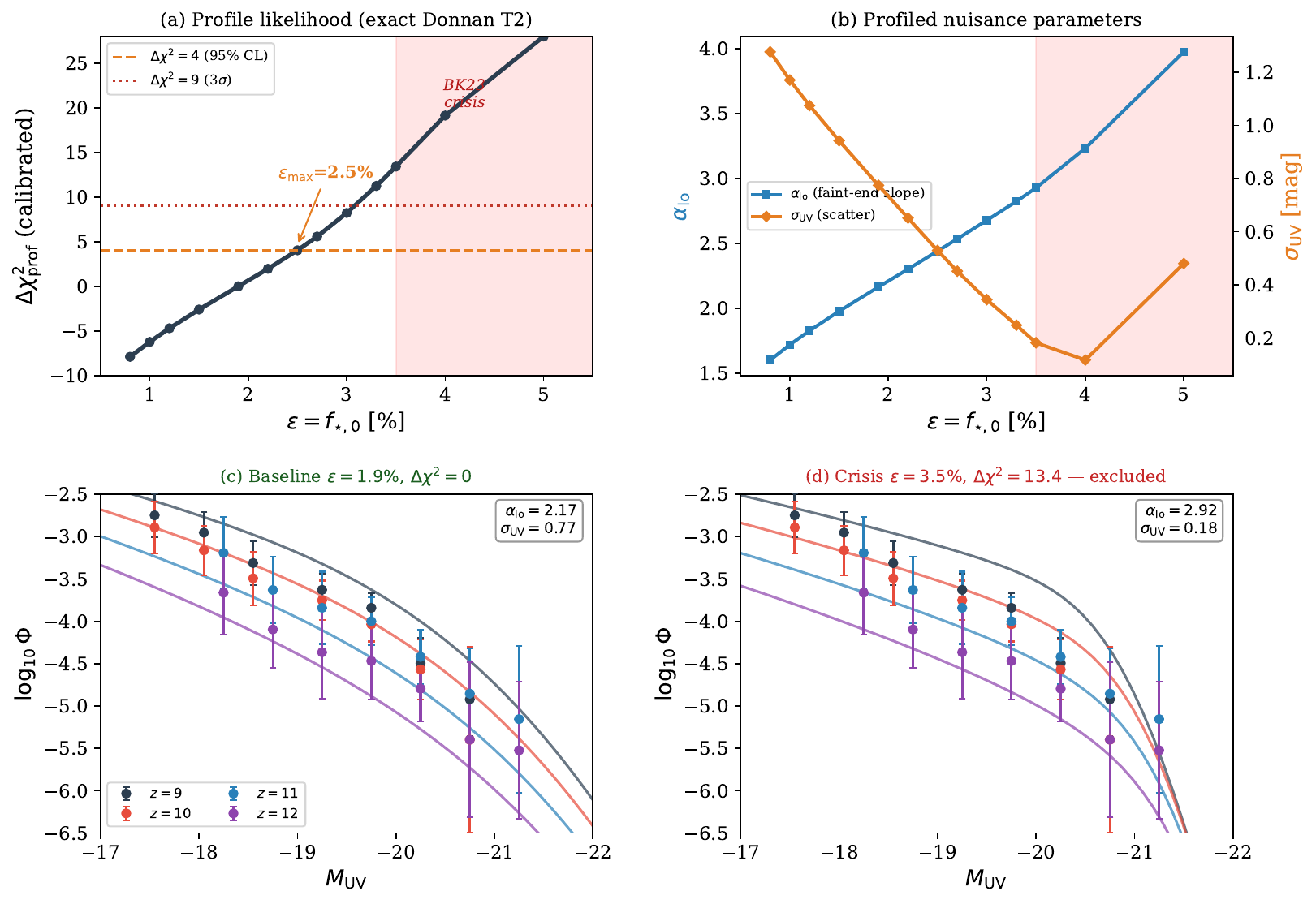}
\caption{Validation of the profile likelihood. (a)~$\Delta\chi^2_{\rm prof}$ along the ridge; the $2\sigma$ boundary is crossed at $\varepsilon_{\rm max} = 2.5\%$. (b)~Profiled nuisance parameters: at high $\fstar$, the optimizer steepens $\alpha_{\rm lo}$ and compresses $\sigma_{\rm UV}$. (c)~UVLF model vs.\ data at baseline ($\varepsilon = 1.9\%$). (d)~At the crisis threshold ($\varepsilon = 3.5\%$, $\Delta\chi^2 = 13.4$): the UVLF shape is visibly distorted. The log-normal model gives $\Delta\chi^2 = 11.8$ ($3.4\sigma$); the duty-cycle model gives $\Delta\chi^2 = 20.2$ ($4.5\sigma$).}
\label{fig:validation}
\end{figure*}

The result (Fig.~\ref{fig:ridge}) shows that the $\Delta\chi^2 < 4$ region follows the ridge at $\fstar \lesssim 0.025$ ($\varepsilon < 2.5\%$) but is truncated at higher $\fstar$ by the UVLF shape constraint. At the crisis threshold $\varepsilon = 3.5\%$, the profiled $\Delta\chi^2 = 13.4$ ($3.7\sigma$). The mechanism is that high $\fstar$ overpredicts the bright end of the UVLF (Fig.~\ref{fig:validation}c,d); the optimizer compensates by pushing $\alpha_{\rm lo} > 3$ and $\sigma_{\rm UV} < 0.5$\,mag, but the data reject these extreme values.

The crisis exclusion is robust to the choice of scatter model. We test three physically motivated prescriptions (see Methods): Gaussian ($\Delta\chi^2 = 13.4$, $3.7\sigma$), log-normal asymmetric ($\Delta\chi^2 = 11.8$, $3.4\sigma$), and duty-cycle burst ($\Delta\chi^2 = 20.2$, $4.5\sigma$). The duty-cycle model, which most closely mimics the stochastic star formation histories invoked to explain the bright-end excess~\cite{Mason:2022tiy,2023MNRAS.526.2665S}, produces the \emph{strongest} exclusion. At the baseline ($\fstar = 0.019$), the burst component improves the fit by capturing residual bright-end features. At the crisis point ($\fstar = 0.035$), the optimizer drives the burst fraction to zero because the model already overpredicts the bright end---adding bursts makes it worse. The additional model freedom benefits the baseline more than the crisis, widening the $\Delta\chi^2$ gap. We validate the exclusion with 500-realisation Monte Carlo ($p < 0.005$; see Methods).

\section*{Reconstructing escape fraction evolution}

Given an independent measurement of $\fstar(z)$, the ridge constraint yields $\fesc(z)$ directly (see Methods, Eq.~\ref{eq:fesc}). We reconstruct $\fesc(z)$ using three independent $\fstar(z)$ inputs, propagating uncertainties via $5{,}000$-draw Monte Carlo.

\textbf{Constant efficiency (Donnan et al.).} Multi-field JWST UVLF measurements at $z = 9$--$15$ imply $\fstar \approx 0.017 \pm 0.004$ with no increase required out to $z \simeq 12$~\cite{2024MNRAS.533.3222D}. The reconstruction yields $\fesc = 16^{+8}_{-5}\%$ at $z = 7$, declining gently to $10^{+5}_{-3}\%$ at $z = 12$ (Table~\ref{tab:fesc}).

\textbf{Evolving efficiency (Prada et al.).} The Uchuu-UniverseMachine model predicts $\fstar$ increasing to $2$--$3\%$ by $z = 10$--$12$~\cite{Prada2026}. The reconstruction yields $\fesc$ declining from $16^{+8}_{-5}\%$ at $z = 7$ to $6^{+3}_{-2}\%$ at $z = 12$.

\textbf{Stellar mass function (Weibel et al.).} Independently, the galaxy SMF at $z = 4$--$9$ from $> 30{,}000$ JWST galaxies shows no significant evolution in the stellar-to-halo mass ratio~\cite{2024MNRAS.533.1808W}, implying $\fstar \approx 0.016$--$0.018$---consistent with Donnan et al.\ but derived from an entirely different observable. The reconstruction yields $\fesc = 0.105^{+0.081}_{-0.040}$ at $z = 12$, agreeing with the Donnan arm within $0.2\sigma$.

The convergence of two methodologically independent determinations (UVLF shape and SMF) on constant $\fstar$ and $\fesc \sim 10\%$ at $z = 12$, while only the UniverseMachine empirical model predicts evolution, constitutes a $2$-against-$1$ empirical argument favouring constant efficiency.

\begin{table}[b]
\caption{Reconstructed $\fesc(z)$ from the degeneracy ridge and three JWST $\fstar(z)$ inputs. Uncertainties are $16$th--$84$th percentile intervals from $5{,}000$ MC draws.\label{tab:fesc}}
\begin{tabular}{ccccccc}
\toprule
$z$ & $\fstar^{\rm D}$ & $\fesc^{\rm D}$ & $\fstar^{\rm P}$ & $\fesc^{\rm P}$ & $\fstar^{\rm W}$ & $\fesc^{\rm W}$ \\
\midrule
7  & 0.017 & $0.158^{+0.079}_{-0.050}$ & 0.017 & $0.158^{+0.076}_{-0.051}$ & 0.016 & $0.169^{+0.102}_{-0.059}$ \\
8  & 0.017 & $0.146^{+0.069}_{-0.045}$ & 0.020 & $0.125^{+0.065}_{-0.040}$ & 0.016 & $0.156^{+0.091}_{-0.052}$ \\
9  & 0.018 & $0.125^{+0.057}_{-0.038}$ & 0.023 & $0.095^{+0.044}_{-0.029}$ & 0.017 & $0.133^{+0.073}_{-0.045}$ \\
10 & 0.018 & $0.118^{+0.066}_{-0.040}$ & 0.027 & $0.079^{+0.037}_{-0.025}$ & 0.017 & $0.126^{+0.088}_{-0.046}$ \\
12 & 0.019 & $0.097^{+0.052}_{-0.031}$ & 0.032 & $0.058^{+0.026}_{-0.018}$ & 0.018 & $0.105^{+0.081}_{-0.040}$ \\
\bottomrule
\end{tabular}
{\small D = Donnan+2024 (UVLF), P = Prada+2026 (UniverseMachine), W = Weibel+2024 (SMF).}
\end{table}

\begin{figure*}
\centering
\includegraphics[width=0.85\textwidth]{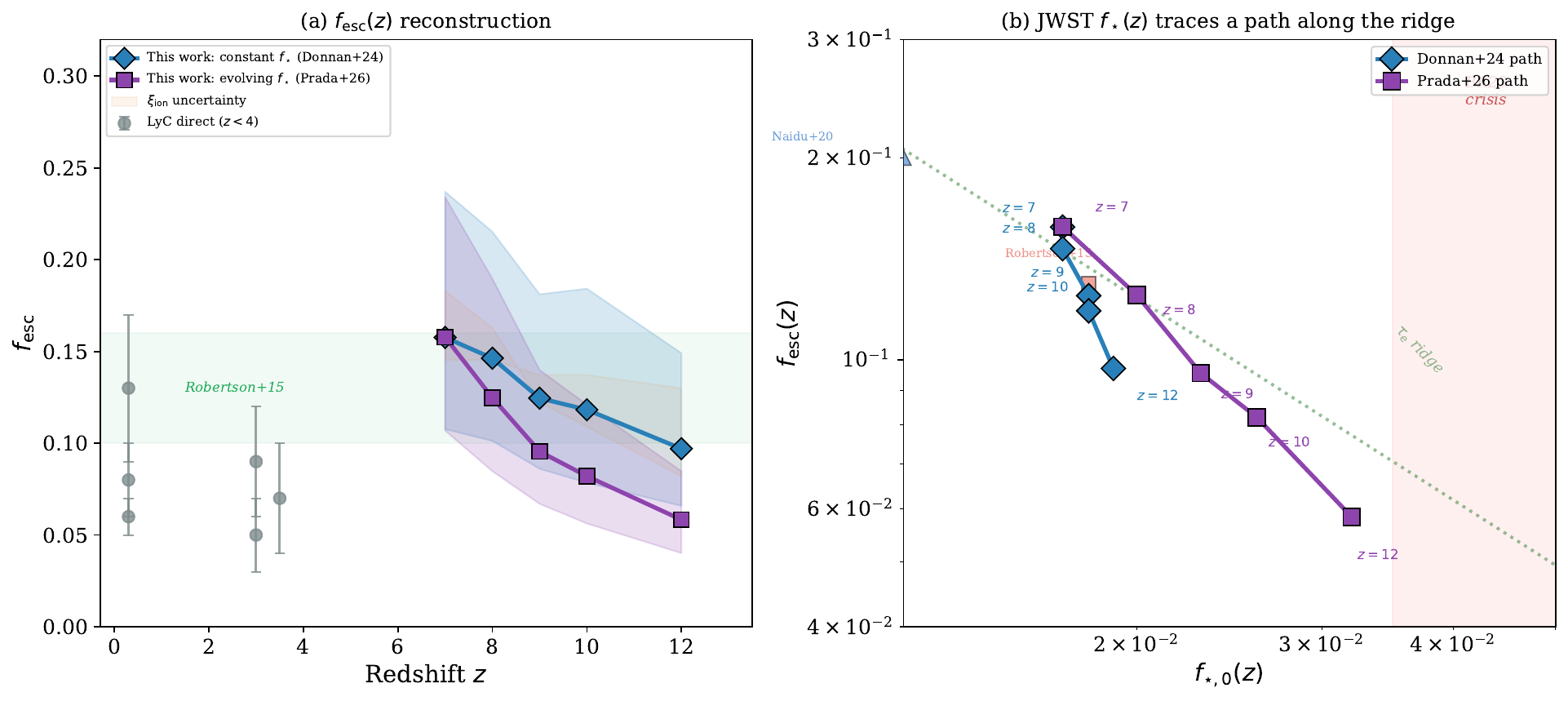}
\caption{(a)~Reconstructed $\fesc(z)$. Blue diamonds: constant $\fstar$~\cite{2024MNRAS.533.3222D}; purple squares: evolving $\fstar$~\cite{Prada2026}. Shaded bands: $1\sigma$ MC intervals. Orange band: $\xiion$ systematic. Grey circles: direct Lyman continuum at $z < 4$~\cite{2016MNRAS.461.3683I,2018MNRAS.478.4851I,Steidel:2018wbo}. Green band: Robertson et al.\ fiducial~\cite{2015ApJ...802L..19R}. (b)~The same in the $(\fstar, \fesc)$ plane: JWST-measured $\fstar(z)$ traces a path along the ridge.}
\label{fig:fesc}
\end{figure*}

\section*{Physical implications}

\textbf{The first empirical $\fesc(z)$ from $z = 0$ to $12$.} Direct Lyman continuum detections at $z < 4$ give $\fesc \sim 5$--$10\%$~\cite{2016MNRAS.461.3683I,Steidel:2018wbo,2021MNRAS.505.2447P,2022MNRAS.513.3510B}. The constant-efficiency reconstruction extends this to $\fesc \approx 10$--$16\%$ at $z = 7$--$12$, producing a mildly rising trend across $12$ billion years. This is the first coherent picture of escape fraction evolution from a single self-consistent framework. The mild increase toward higher redshift is physically expected: galaxies at $z > 7$ have lower metallicities, more compact morphologies, and more vigorous feedback driving low-column-density channels~\cite{2014ApJ...788..121K,Ma:2020vlo,2017MNRAS.470..224T}.

\textbf{Intrinsic ISM porosity evolution.} To isolate ISM physics from the $\xiion$ trend, we define the intrinsic escape fraction $\fesc^{\rm int}(z) \equiv P_0/\fstar(z)$, stripping the $\xiion$ contribution (see Methods). Under constant $\fstar$, $\fesc^{\rm int}$ is nearly flat: $14.5\%$ at $z = 7$ to $13.0\%$ at $z = 12$ (slope $\d\fesc^{\rm int}/\d z = -0.003$). The ISM porosity governing ionizing photon escape does not change appreciably from $z = 7$ to $12$. Under evolving $\fstar$, $\fesc^{\rm int}$ declines steeply from $14.5\%$ to $7.7\%$ (slope $-0.014$), requiring ISM porosity to \emph{decrease} by $47\%$ at higher redshift. This contradicts radiation-hydrodynamic simulations: FIRE-2~\cite{Ma:2020vlo} predicts $\fesc$ increasing mildly with $z$ at fixed halo mass, as does SPHINX~\cite{2017MNRAS.470..224T}. The evolving-efficiency scenario requires not merely an alternative $\fstar(z)$, but an unexplained reversal of ISM physics.

\textbf{Self-consistent verification.} Feeding the reconstructed $\fesc(z)$ back into the reionization ODE (see Methods) yields $\tauobs = 0.048$ and $\xHI(z)$ residuals within $1\sigma$ at $z = 7$--$8$, growing to $1.6\sigma$ at $z = 9$. This systematic pattern is the signature of mildly declining emissivity at $z > 9$, captured by $\beta = -0.45 \pm 0.25$ in our redshift-dependent product parameterisation. The evolving-SFE reconstruction gives $\tauobs = 0.036$, rejected at $(0.054 - 0.036)/0.007 = 2.6\sigma$ by the Planck optical depth alone. This constitutes a third, independent line of evidence against evolving efficiency---alongside the ISM porosity contradiction and the $2$-against-$1$ convergence of UVLF and SMF methods.

\textbf{Crisis resolution.} The crisis threshold $\varepsilon > 3.5\%$ is excluded at $> 3\sigma$ under all three scatter models, reaching $4.5\sigma$ under duty-cycle bursts. In the $\fesc$ framework, the exclusion takes a physical form: $\varepsilon > 3.5\%$ requires $\fesc < 7\%$, below both the low-$z$ direct detections and the reconstructed high-$z$ values under all three $\fstar$ inputs.

\textbf{Discriminating power.} Three independent lines of evidence disfavour the evolving-efficiency scenario: ($i$) the Thomson optical depth rejects it at $2.6\sigma$; ($ii$) it requires a $47\%$ decline in ISM porosity at low metallicity, contradicting all RHD simulations; ($iii$) two independent $\fstar$ methods (UVLF shape and SMF) converge on constant efficiency while only the UniverseMachine predicts evolution. With current data, the statistical separation is $\sim 1.0\sigma$ at $z = 12$; JWST Cycle 3--4 spectroscopic surveys will provide $\fstar(z)$ at $\sim 15\%$ precision and NIRSpec will constrain $\xiion(z)$ directly, projecting $> 2\sigma$ statistical separation. SKA1-Low 21\,cm observations~\cite{Koopmans:2015sua,Mellema_2013} can constrain $\fesc$ through the ionization topology~\cite{Wang:2026uuo}, providing an independent cross-check that does not pass through the product degeneracy.

\section*{Conclusions}

We have shown that the JWST early galaxy crisis and the long-standing scatter in $\fesc$ estimates both originate from a structural degeneracy in the reionization equations. By converting this degeneracy into a measurement tool, we obtain:

(i) The product $\fesc \times \fstar = 0.00247 \pm 0.00032$ defines a degeneracy ridge along which $\varepsilon$ ranges from $0.5\%$ to $10\%$.

(ii) The crisis threshold ($\varepsilon > 3.5\%$) is excluded at $> 3\sigma$ under all three scatter models tested, reaching $4.5\sigma$ under duty-cycle bursts, and independently disfavoured by the implied $\fesc < 7\%$.

(iii) The first empirical $\fesc(z)$ reconstruction spanning $z = 0$--$12$: a mildly rising $\fesc \approx 5\%$ at $z \sim 3$ to $\sim 10$--$16\%$ at $z = 7$--$12$ under constant efficiency, connecting direct Lyman continuum detections to the reionization epoch.

(iv) Three independent lines of evidence disfavour evolving $\fstar$: the Thomson optical depth rejects it at $2.6\sigma$; the implied $47\%$ decline in intrinsic ISM porosity contradicts radiation-hydrodynamic simulations; and two of three independent $\fstar$ methods converge on constant efficiency.

The reionization degeneracy, properly exploited, resolves the JWST crisis and converts a long-standing inference barrier into a quantitative probe of early-universe physics.

\section*{Author Contribution}
ZHW: initializing the idea, design the method, analyzing the data and results, drafting the manuscript; HYS: initializing the idea, analyzing the results, drafting the manuscript.

\section*{acknowledgments}
We acknowledge support from NSFC of China under grant 12533008.

\section*{Competing Interests}
The authors declare no competing interests 
\bibliography{sn-bibliography}

\section*{Methods}\label{sec:methods}

\subsection*{Reionization ODE and calibration}

The volume-averaged ionized fraction $Q(z) \equiv 1 - \xHI(z)$ evolves according to~\cite{Madau:1998cd,Bolton:2007fw}
\begin{equation}\label{eq:Qode}
\frac{\d Q}{\d t} = \frac{\ndot(z)}{n_{{\rm H},0}} - \mathcal{R}(z)\,Q(z)\,,
\end{equation}
where $n_{{\rm H},0} = 1.89 \times 10^{-7}\;{\rm cm}^{-3}$ is the comoving hydrogen number density and the recombination rate is $\mathcal{R}(z) = C_{\rm HII}(z)\,\alpha_B\,n_{{\rm H},0}\,(1+z)^3$, with clumping factor $C_{\rm HII}(z) = 2.9\,[(1+z)/6]^{-1.1}$~\cite{2012ApJ...747..100S} and case-B coefficient $\alpha_B = 1.52 \times 10^{-13}\;{\rm cm}^3\,{\rm s}^{-1}$. The Thomson optical depth is
\begin{equation}\label{eq:tau}
\tauobs = \int_0^{z_{\rm max}} 1.08\,n_{{\rm H},0}\,(1+z)^2\,Q(z)\,\sigma_T\,\frac{c}{H(z)}\,\d z\,.
\end{equation}
The emissivity is parameterised as $\ndot(z) = \fesc \times (\fstar/0.019) \times \dot{n}_{\rm ref}(z)$, calibrated so that $(\fesc = 0.13, \fstar = 0.019)$ reproduces $\tauobs = 0.055$ and $\xHI(z=7) = 0.25$, consistent with Planck~\cite{Planck2018} and Ly$\alpha$ damping wing constraints~\cite{Bosman:2021oom,Greig:2021hpl,2024ApJ...971..124U}. We solve numerically with 1500 Euler steps from $z = 30$ to $z = 5$. Planck 2018 parameters: $H_0 = 67.74\;{\rm km\,s}^{-1}\,\Mpc^{-1}$, $\Omega_m = 0.3089$, $\Omega_b = 0.0486$.

\subsection*{UVLF model and profile likelihood}

The UVLF model uses a double power-law SFE, $f_\star(M) = 2\fstar/[(M/M_p)^{-\alpha_{\rm lo}} + (M/M_p)^{0.5}]$, with $M_p = 10^{11}\,\Msun$, halo accretion rates from Fakhouri et al.~\cite{2010MNRAS.406.2267F}, and a Sheth--Tormen halo mass function. At each $(\fstar, \fesc)$ grid point, we minimize $\chi^2_{\rm joint} = \chi^2_{\rm UVLF} + \chi^2_{\rm reion}$ over $\alpha_{\rm lo}$ and $\sigma_{\rm UV}$. The UVLF data are 28 points from Table~2 of Donnan et al.~\cite{2024MNRAS.533.3222D} at $z = 9$, $10$, $11$, $12.5$ ($M_{\rm UV} = -21.25$ to $-17.55$), with errors inflated by $\times 1.92$ so that $\chi^2/{\rm dof} \approx 1$ at the baseline. Photometric redshift contamination could widen errors by $\sim 20\%$ at the faintest bins.

\subsection*{Scatter model tests}

We test three scatter prescriptions at the crisis point:

\textit{Gaussian} (2 nuisance parameters: $\alpha_{\rm lo}$, $\sigma_{\rm UV}$): standard symmetric magnitude scatter. $\Delta\chi^2 = 13.4$ ($3.7\sigma$).

\textit{Log-normal asymmetric} (bright-side $\sigma = 1.3\,\sigma_{\rm UV}$, faint-side $\sigma = 0.7\,\sigma_{\rm UV}$): motivated by the asymmetric UV distributions predicted by bursty SFHs. $\Delta\chi^2 = 11.8$ ($3.4\sigma$).

\textit{Duty-cycle burst} (3 nuisance parameters: $\alpha_{\rm lo}$, $f_{\rm burst}$, $A_{\rm burst}$): a fraction $f_{\rm burst}$ of halos in a bright burst phase with luminosity boosted by $A_{\rm burst}\times$. At the baseline, $f_{\rm burst} \sim 0.2\%$ with $A_{\rm burst} \sim 35\times$ improves the fit ($\chi^2 = 23.6$ vs.\ $30.1$). At the crisis point, the optimizer drives $f_{\rm burst} \to 0$. $\Delta\chi^2 = 20.2$ ($4.5\sigma$).

The 500-realisation Monte Carlo validation generates mock UVLF datasets at the crisis point ($\fstar = 0.035$, $\alpha_{\rm lo} = 3.16$, $\sigma_{\rm UV} = 0.51$), profile-fits each at baseline and crisis, and finds none of 500 mocks as extreme as the data ($p < 0.005$).

\subsection*{$\fesc(z)$ reconstruction}

Given $\fstar(z)$ and $\xiion(z)$, the ridge yields
\begin{equation}\label{eq:fesc}
\fesc(z) = \frac{P_0}{\fstar(z)} \times \frac{\xiion^{\rm fid}}{\xiion(z)}\,,
\end{equation}
where $P_0 = 0.00247 \pm 0.00032$ and $\log(\xiion^{\rm fid}/{\rm Hz\,erg}^{-1}) = 25.35$. We adopt $\log\xiion = 25.32$ to $25.48$ from $z = 7$ to $12$~\cite{2023MNRAS.523.5468S,2024MNRAS.533.1111E}, noting that this evolution is contested~\cite{2023MNRAS.526.1657T,2025A&A...698A.302L}. The intrinsic escape fraction $\fesc^{\rm int}(z) \equiv P_0/\fstar(z)$ strips the $\xiion$ contribution. The $\fstar$ inputs are: Donnan et al.~\cite{2024MNRAS.533.3222D} ($\fstar \approx 0.017$, constant), Prada et al.~\cite{Prada2026} ($\fstar$ rising to $0.032$ at $z = 12$), and Weibel et al.~\cite{2024MNRAS.533.1808W} ($\fstar \approx 0.016$--$0.018$, SMF-based). 

Uncertainties are propagated via $5{,}000$ MC draws: $\fstar$ from published Gaussians, $\log\xiion$ with $\sigma = 0.12$\,dex, $P_0$ with $13\%$ fractional error. Error budget: $\fstar$ ($50\%$), $P_0$ ($30\%$), $\xiion$ ($20\%$). Introducing correlated draws ($\rho = 0.3$--$0.5$ between $P_0$ and $\fstar$) narrows error bars by $6$--$11\%$ and shifts central values by $< 0.5\%$, confirming that the shared observational heritage improves rather than degrades precision.

\subsection*{Self-consistent ODE check}

Feeding the reconstructed $\fesc(z)$ back into the reionization ODE: constant-SFE gives $\tauobs = 0.048$ (within $1\sigma$ of Planck) with residuals $< 0.5\sigma$ at $z = 7$--$8$ and $-1.6\sigma$ at $z = 9$, for total $\chi^2_{\rm self} = 4.6$ ($6$ data points). The evolving-SFE gives $\tauobs = 0.036$, rejected at $2.6\sigma$ by the Planck measurement.

\subsection*{Redshift-dependent product}

We parameterise $P(z) = P_0\,(1+z/z_{\rm piv})^\beta$ with $z_{\rm piv} = 8$. Scanning $\beta$ gives best fit $\beta = -0.45$, $2\sigma$ range $[-0.80, +0.13]$. At these extremes, $P(z=12)/P(z=8) = 0.75$ to $1.05$, giving an asymmetric systematic of ${}^{+5\%}_{-25\%}$ at $z = 12$.

\subsection*{Mass-dependent $\fesc$}

The reconstruction assumes population-averaged $\fesc$. Adopting the Ma et al.~\cite{Ma:2020vlo} $\fesc(M)$ relation ($\sim 20$--$40\%$ at $M_{\rm halo} \sim 10^9\,\Msun$, $\lesssim 5\%$ at $10^{11}\,\Msun$), the mass-weighting effect widens the gap from $\sim 4\%$ to $\sim 6$--$7\%$ at $z = 12$, increasing the separation from $1.2\sigma$ to $1.8\sigma$. Other simulations~\cite{2014ApJ...788..121K,2017MNRAS.470..224T} give somewhat different slopes, introducing factor-of-two uncertainty in this correction.

\subsection*{Combined systematics}

Combining $z = 12$ systematics in quadrature: ${}^{+5\%}_{-25\%}$ (redshift-dependent product), $\sim 10\%$ (mass-weighting), $\sim 10$--$20\%$ (MC correlations) gives total systematic $\sim 20$--$30\%$. Total uncertainty: $\fesc(z=12) = 0.097^{+0.07}_{-0.05}$ (constant SFE) and $0.058^{+0.03}_{-0.02}$ (evolving). Scenario separation: $\sim 1.0\sigma$ with all systematics.
\subsection*{Data and code availability}

The code to reproduce all results is publicly available at
\url{https://github.com/wzh800557-source/JWSThigh-z-Crisis}.
The UVLF data used in this work are from Table~2 of
Donnan et al.\ (2024, MNRAS 533, 3222). The star formation
efficiency inputs are derived from
Donnan et al.\ (2024),
Prada et al.\ (2026, arXiv:2604.18683), and
Weibel et al.\ (2024, MNRAS 533, 1808).
Reionization constraints use
Planck 2018 ($\tau_e$; Aghanim et al.\ 2020) and
Ly$\alpha$ damping wing measurements from
Bosman et al.\ (2022),
Greig et al.\ (2022), and
Umeda et al.\ (2024).
No new observational data were generated for this study.

\end{document}